\documentclass[conference]{IEEEtran}
\usepackage{ graphicx}

\usepackage{amssymb}
\usepackage{amsthm}
\usepackage{lineno}
\usepackage{graphicx}  %Required
\graphicspath{ {images/} }
\usepackage{cite}
\usepackage{graphicx}
\usepackage{subcaption}
\usepackage[utf8]{inputenc}
\usepackage[export]{adjustbox}
\usepackage{wrapfig}
\usepackage{array}
\usepackage[normalem]{ulem}
\usepackage{adjustbox}
\usepackage{fancybox}
\usepackage{array}
\usepackage{booktabs}
\usepackage{multirow}
\usepackage{array}
\usepackage{xspace}
\usepackage{cleveref}
\usepackage{longtable}
\graphicspath{ {images/} }
\usepackage{subcaption}
\usepackage{amsmath}
\usepackage{supertabular}
\usepackage{url}
\usepackage{bm}
\usepackage{amssymb}
\usepackage{amsmath}
\usepackage{wrapfig}
\usepackage{amsmath}
\usepackage{longtable}
\usepackage[linesnumbered,ruled,vlined]{algorithm2e}

\usepackage{algorithmic}
    
\graphicspath{ {images/} }
\ifCLASSINFOpdf
\else
\fi

\begin{document}
\title{Enabling Trust in Deep Learning Models: \\ A Digital Forensics Case Study  \\
}

\author{
\IEEEauthorblockN{Aditya. K. }
\IEEEauthorblockA{School of Computer Science\\
University College, Dublin \\
Email: aditya.kuppa@ucdconnect.ie}
\and
\IEEEauthorblockN{Slawomir Grzonkowski }
\IEEEauthorblockA{
Email: slawomir\_grzonkowski@symantec.com }
\and
\IEEEauthorblockN{NhienAn LEKHAC}
\IEEEauthorblockA{School of Computer Science\\
University College, Dublin \\
Email: an.lekhac@ucd.ie  } }

\newcommand{\prname}{ATF\xspace}
\newcommand{\ie}{i.e., }
\newcommand{\eg}{e.g., }

\newcommand{\stitle}[1]{\vspace{5pt} \noindent\textbf{#1.}\ }
\newcommand{\todo}[1]{{\color{blue} {\bf TODO:} {\small #1}}}
\newtheorem{theorem}{Theorem}
\newtheorem{lemma}{Lemma}
\newtheorem{proposition}{Proposition}
\newtheorem{remark}{Remark}
\newtheorem{definition}{Definition}

\newcommand{\esm}[1]{\ensuremath{#1}}
\newcommand{\mr}[1]{\esm{\mathrm{#1}}}
\newcommand{\ms}[1]{\esm{\mathsf{#1}}}
\newcommand{\mi}[1]{\esm{\mathit{#1}}}
\newcommand{\mb}[1]{\esm{\mathbf{#1}}}
\newcommand{\mathsc}[1]{{\normalfont \textsc{#1}}}
\newcommand{\msc}[1]{\esm{\mathsc{#1}}}

\newcommand\reals{\ms{R}} \newcommand\incp{\ms{Incp}}
\newcommand\maxf{\ms{max}} \newcommand\img{\ms{img}}
\newcommand\pimportance{\mathcal{P}}
\newcommand\molfeatures{\ms{Features}}

\newcommand\grad{\bigtriangledown}
\newcommand\sparam{\alpha}
\newcommand\im{\ms{img}}
\newcommand\mol{\ms{mol}}
\newcommand\pathfn{\gamma}

\newcommand\synteq{::=}
\newcommand\integratedgrads{\ms{IG}}

\newcommand\intgrads{\ms{InteriorGrads}}
\newcommand\pathintegratedgrads{\ms{PathIntegratedGrads}}
\newcommand\relu{\ms{ReLU}}
\newcommand\sigmoid{\ms{Sigmoid}}
\newcommand\xbase{x'}
\newcommand{\fix}{\marginpar{FIX}} \newcommand{\new}{\marginpar{NEW}}

\newcommand\intpimportance{\ms{InteriorPixelImportance}}

\newcommand*{\myfont}{\fontfamily{phv}\selectfont}
\newcommand{\V}[1]{\mathbf{#1}}
\newcommand{\bfC}{\V{C}}
\newcommand{\bfu}{\V{u}}
\newcommand{\bfU}{\V{U}}
\newcommand{\bfw}{\V{w}}
\newcommand{\bfv}{\V{v}}
\newcommand{\bfX}{\V{X}}
\newcommand{\bfx}{\V{x}}
\newcommand{\bfy}{\V{y}}
\newcommand{\xad}{\bfx_{\text{adv}}}
\newcommand{\Xad}{\bfX_{\text{adv}}}
\newcommand{\bfZ}{\bm{\phi}}
\newcommand{\Attack}[1]{\mathcal{A}\left(#1\right)}
\newcommand{\tran}{^{\mkern-1.5mu\mathsf{T}\mkern-1.5mu}}

\maketitle

\begin{abstract}

  Today, the volume of evidence collected per case is growing exponentially, to address this problem forensics investigators are looking for investigation process with tools built on new technologies like big data, cloud services and Deep Learning (DL) techniques. Consequently, the accuracy of artifacts found also relies on the performance of techniques used, especially DL models. Recently, \textbf{D}eep \textbf{N}eural \textbf{N}ets (\textbf{DNN}) have achieved state of the art performance in the tasks of classification and recognition. In the context of digital forensics, DNN has been applied in the domains of cybercrime investigation such as child abuse investigations, malware classification, steganalysis and image forensics. However, the robustness of DNN models in the context of digital forensics is never studied before. Hence, in this research, we design and implement a domain independent Adversary Testing Framework (ATF) to test security robustness of black-box DNN's. By using ATF, we also methodically test a  commercially available DNN service used in forensic investigations and bypass the detection, where published methods fail in control settings.

\end{abstract}


\IEEEpeerreviewmaketitle

\section{Introduction}
Broadly, digital forensics investigation process consists of - Identify, collect and acquire evidence, examining and analyzing the collected evidence and finally, document and report observations without compromising data integrity. Today, the amount of material presented for digital forensic examination has increased significantly, this has left many police forces and organizations facing a large backlog of examination\cite{Ben16}. In addition, due to the availability of cheap storage, increasing adoption of IPv6, using of  cloud services, and increased use of mobile devices for a single case, forensic data today becomes Big Data\footnote{http://www.gartner.com/it-glossary/big-data} in the Internet of Things(IoT) environment\cite{Saad18}. Hence, recently digital investigators have applied big data technology, cloud-based forensics services, as well as Machine Learning (ML) and deep Learning techniques  to collect and analyze efficiently a large volume of heterogeneous evidence\cite{Choo17}.

In fact, machine learning algorithms such as logistics regression, support vector machines, decision trees have been used for malware classification and forensics artifact correlation\cite{for_arc}\cite{Andree16}. In audio forensics, unsupervised learning techniques have been applied to separate voice from background noise. Clustering, outlier detection, and dimensionality reduction methods are used for analyzing of network data\cite{2017arXiv170906599U}\cite{Loic16} and financial fraud analysis\cite{van14}\cite{an2009}. 

The recent success of Deep Neural Nets (DNN) in the tasks of classification and recognition have led to widespread adoption and deployment of DNN in security and safety-critical systems. It is no longer a promising but immature technology, as DNN systems have reached close to human-level performance on unseen data. In the context of digital forensics DNN is proposed for steganalysis\cite{stegno}, video forgery detection\cite{video-forgery}, image forensics\footnote{https://www.guidancesoftware.com/app/Image-Analyzer-Day-Free-Trial} and commercially used in digital forensic tools.

Security related applications based on ML and DNN models have to take into account adversarial settings and scenarios. An intelligent and adaptive adversary can submit a carefully crafted input to a classification system to purposely undermine its performance. It is now acknowledged that DNN exhibit multiple vulnerabilities and are prone to various attacks both targeted and untargeted\cite{2018arXiv180100553}. In order to allow successful adaption and trusting of  DNN outputs in digital forensics settings, there is a need for systematic evaluation of the robustness of DNN models used in forensics tools. However, to the best of our knowledge, there is no existing study which tests the reliability, robustness, and stability of digital forensics tool which uses DNN models in the literature.

The key challenges of testing security robustness of black-box deep learning systems are two-fold: (1) how to generate inputs which uncover different erroneous behaviors, and (2) how to adapt black-box attacks to different threat models and adversary resistant settings. Hence, this paper describes how we design and build  Adversary Testing Framework (ATF)  to address both challenges. In summary, our contributions include: 

\begin{itemize}
  \item We extend the taxonomy of threat models for testing security robustness of black-box DNN.
  \item We define standard empirical metrics to measure the effectiveness of attacks on black-box DNN.
  \item We design and implement Adversary Testing Framework (ATF) for testing robustness of real-world hosted black box DNN models.
  \item We carry out the practical black-box attack using ATF  against Digital Forensics API - Image Analyzer and show that these models can be bypassed with high confidence.
  \end{itemize}

\section{Preliminaries}
\label{prem}
Multiple attacks have been demonstrated against DNN models in controlled settings. Recent work also illustrates the possibility of attacks on deployed systems such as  ML as a service cloud applications \cite{papernot2016transferability} in laboratory settings. Latest survey \cite{2018arXiv180100553} summarizes different attacks published  in the literature and Table \ref{tab:2} captures different attacks on DNN's in non-image domains.

A  qualitative taxonomy for the threat models\cite{Barreno_2006,Huang_2011,Papernot_distill_2016} against machine learning systems  places the attacks in following  axes:
\begin{itemize}

\item \textbf{Influence:} can be either causative or exploratory. A causative 
influence  can affect the learning process, while an exploratory attack only deals with a trained
classifier. Attacks that affect the learning process have also been categorized as
poisoning and attacks that are used only at test time have also been described
as evasion attacks.

\item \textbf{Security Violations:} can be integrity related when the adversary can cause
misclassification of attacks to appear as normal, availability related when
the misclassification are so many that the system becomes unusable and privacy violation, when the adversary obtains
information from the learner,  compromising the secrecy or privacy of the system’s users.

\item \textbf{Specificity:} can be targeted when it focuses on specific misclassification or
indiscriminate where it does not matter which classes are being misclassified.

\item \textbf{Complexity of the attack:} which ranges from simple confidence reduction to
complete source/target misclassification.
\item \textbf{Knowledge of the attacker:} which ranges from knowledge about architecture,
training tools and data to just knowledge of a few samples.
If the attacker knows anything regarding the architecture, the training data or the features used, the attack is considered a white-box attack. If the adversary's knowledge is limited to Oracle attacks or she has only access to a limited number of samples, the
attack is considered a black-box attack.
\end{itemize}

In a security-sensitive setting, attacker goal is not only to compromise security of the system but also stay undetected i.e. not raise any alerts or errors in the underlying system. We propose following \textbf{additional} goals for an attacker when designing  threat model for a security sensitive black-box DNN:

\begin{itemize}
\item \textbf{Error Sensitivity} - Attacker has to avoid any error raised by underlying detection system.  
\item \textbf{Alert suppressing} - Attacker should not raise an alert in the system which is under attack. 
\item \textbf{Query rate Limiting} - Adversary has to respect query limit imposed by  API when attacking a  remotely hosted DNN API. 
\item \textbf{Persistence  of attack} - Black-box models can be re-trained with new data and replaced with the new model, a successful attack has to survive retraining and replacement of models.
\item \textbf{Multiple sub models} - In a black-box setting, output to a query can be generated by multiple sub-models, the attacker has to fool all or sub-set of models to achieve his goal.
\end{itemize}

\begin{table}[t!]
\centering
\caption{Summary  of  black box attacks on DNN's in non image domains}
\label{tab:2}
\begin{tabular}{|l||c|c|c|}
\hline
 {\bf Domain}             &   Type  &  Targeted/Non-targeted        \\ \hline\hline
Google Video API's  \cite{GoogleVideoAPI}     &   Black box   &  Targeted         \\ \hline
Face Recognition Systems \cite{FaceCrime}     &   Black box   &  Non-targeted           \\ \hline
Text classification systems \cite{poison} &   Black box   &  Non-targeted           \\ \hline
\end{tabular}
\end{table}

In order to measure  the effectiveness  of  black-box attacks, we use the following  metrics.

\begin{itemize}
\item \noindent \textbf{Attack success rate.}
The  fraction of samples that meets the adversary's goal:
$f(\xad) \neq y$ for un-targeted attacks and $f(\xad)=T$ for targeted attacks with target $T$ \cite{tramer2017ensemble}.
\item \noindent \textbf{Average distortion.}
We also evaluate the average distortion for adversarial examples using average $L_2$ distance between the benign samples and the adversarial ones:
  $\Delta(\Xad,\bfX) = \frac{1}{N} \sum_{i=1}^N \|(\Xad)_i - (\bfX)_i\|_2$%

where $N$ is the number of samples. This metric allows us to compare the average distortion for attacks which achieve similar attack success rates, and therefore infer which one is stealthier.
\item \noindent \textbf{Root Mean Square Deviation.}
Distortion is measured by \emph{root mean square deviation} (RMSD), which is computed as
\begin{align}
 \label{eqn_RMSD}
 dev(x^\star, x) = \sqrt{\sum_i(x_i^\star - x_i)^2/N}
 \end{align}
 where $x^\star$ and $x$ are the vector representations of distorted  image
and the original image  respectively, $N$ is the dimensionality of $x$ and $x^\star$,
and $x_i$ denotes the pixel value of the $i$-th dimension of $x$, which is
within range $[0, 255]$.

\item \noindent \textbf{Number of queries.} Query based black-box attacks make queries to the target model, and
this metric may affect the cost of mounting the attack. This is an important consideration when attacking real-world systems which have costs associated with the number of queries made.
\end{itemize}

\section{ Adversary Testing Framework}
\begin{figure}[h]
\centering
    \includegraphics[width=0.8\columnwidth]{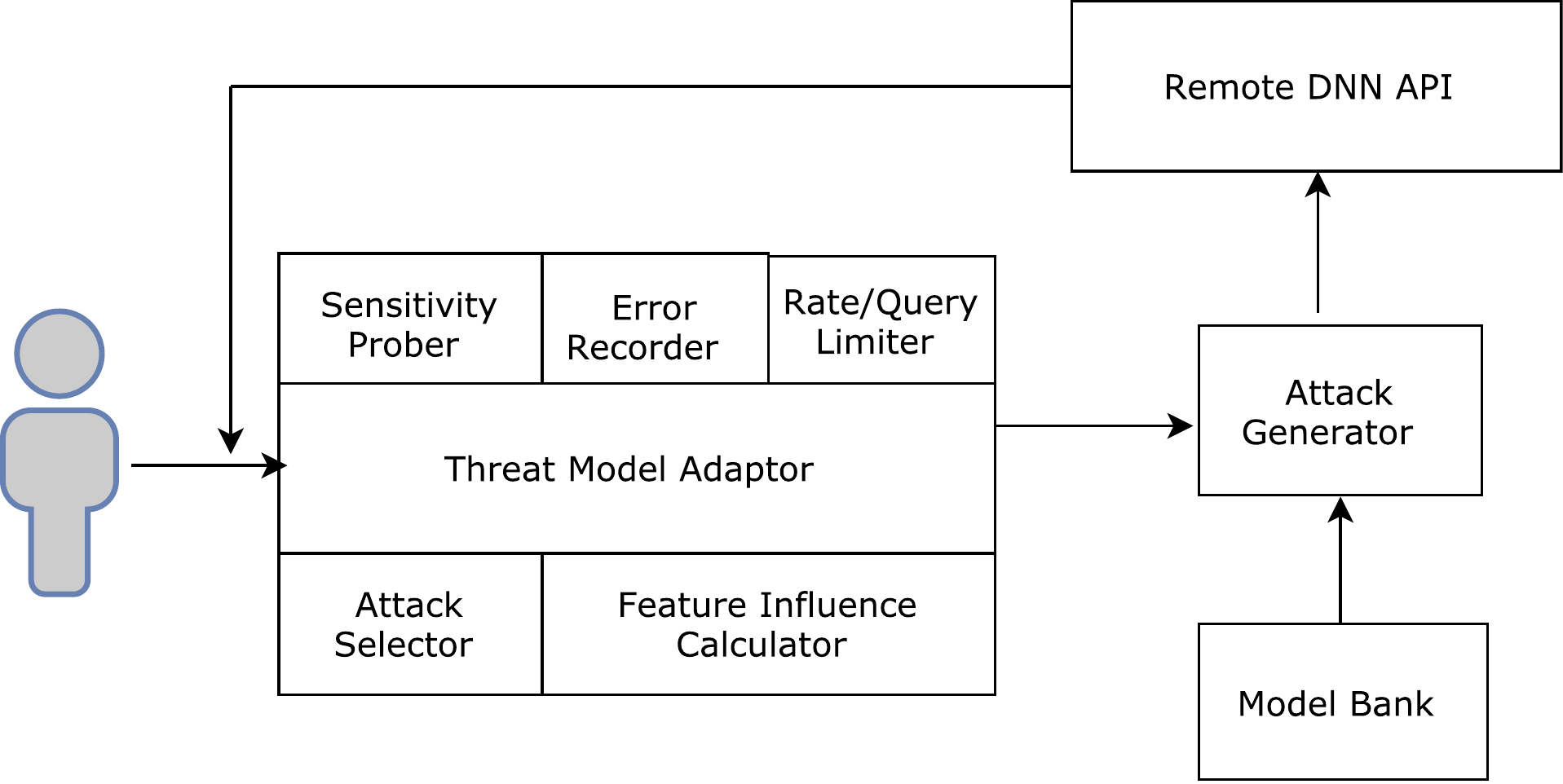}
    \caption{Adversary Testing Framework }
    \label{fig: Framework}   
\end{figure}

In this section, we provide a general overview of, our Adversary  Testing Framework(\prname) for systematically testing black-box DNNs. The main components of are shown in Figure~\ref{fig: Framework}.

\textbf{Threat Model Adapter.}
Threat Model Adapter(TMA) consists of various components which probe the black-box DNN's defenses. Depending on adversary goal, domain, and defenses in place, TMA iteratively crafts attack samples to bypass the security of black-box DNN.
\begin{itemize}

\item \textbf{Feature Influence Calculator(FIC).}
We define Feature Influence as a set of features in input space which are attributed to the prediction of a DNN. We use Integrated Gradients(IG)\cite{ig} technique to calculate feature Influence of remote black-box DNN. Since gradients are not available we use a surrogate model and exploit the transferability to craft the attack samples. IG  is calculated by aggregating the gradients along the inputs that fall on the straight line between the baseline and the input. As shown in equation \ref{ig_eq} integrated gradients method identifies features which define class boundaries for a specific model. The integral of integrated gradients can be efficiently approximated via a summation. We sum the gradients at points occurring at sufficiently small intervals along the straight line path from
the baseline $\xbase$ to the input $x$. For a given input, first, we get feature activations for each model in the model bank. Next, we use query limiter to choose predefined percent of activated features which are found in the majority of models.

\begin{equation}
\label{ig_eq}
  \begin{split}
    \integratedgrads_i^{approx}(x) = 
    (x_i-\xbase_i)\times \Sigma_{k=1}^m  \tfrac{\partial F(\xbase + \tfrac{k}{m}\times(x-\xbase)))}{\partial x_i}\times\tfrac{1}{m}
  \end{split}
\end{equation}

Here $m$ is the number of steps in the Riemman approximation of the
integral.

\item \textbf{Rate/Query Limiter.}
The major drawback of the existing black-box attacks is that the number of queries needed per adversarial sample is large. For an input with dimension $d$, the number of queries will be exactly $2d$. In security-sensitive setting, the number of queries available to an attacker is limited.  The functionality of query limiter is to reduce the number of queries the adversary has to make. We use feature grouping technique to reduce the number of queries. 

In each iteration of Algorithm~\ref{alg: fic_feat}, let there by k models in the model bank. For each  model $m$  we calculate most activated features  $\bfy_m$  for input $\bfx$ using IG, this will  generate  $m$ feature vectors  $\bfy_m$. Now, choose only those features for which multiple models have activated the same feature and cross the threshold $n$. 

\begin{algorithm}
  \caption{Query limiting  using FIC}
    \label{alg: fic_feat}
  \begin{algorithmic}[1]
  \REQUIRE $\bfx$, k, $\bfy$, $n$
  \FOR{$i \gets 1 \textrm{ to } k $}
  \STATE  $\bfy_k$  = IG for $model_k$ for input $\bfx$
  \ENDFOR
    \STATE select features from $\bfy_k$ which are active for count($model_k$) \textgreater  $n$  .   
  \end{algorithmic}
\end{algorithm}

\item \textbf{Sensitivity Prober.}
Prober main functionality is to test black-box model sensitivity towards the random noise, random feature perturbations, and input transformations like feature compression. It also calculates different metrics as described in Section \ref{prem} to generate baselines for the attack. 

\item \textbf{Error Recorder.}
In real life scenario, black-box models generate different types of errors which can be useful to attacker depending on the threat model. Errors can include configurations errors, system generated errors and run-time errors. Errors like system generated and runtime errors may expose underlying defenses used by black-box DNN which can help an attacker to achieve his goal.
The error recorder main function is to record any errors generated by black-box DNN and map them to threat model under test.  

\item \textbf{Attack Selector.}
Given the threat model definition, errors generated by DNN, and sensitivity parameters attack selector selects the best attack suitable for the black-box DNN. It also tunes parameters for a given attack. 

\end{itemize}

\textbf{Model Bank and Attack Generator.}
Ideally, a black-box attack can start with a random input and iteratively run different attacks to discover decision boundaries of underlying black-box DNN. But in real life scenarios, the number of queries available is limited. So, we use an ensemble of models for generating adversarial samples that transfer well to the target model. A collection of DNN models which have different architecture, weights, parameters and follow different training methodology are found in the model bank.
Given a clean input, threat model constraints, attack generator creates adversarial samples to test black-box DNN.

\section{Case Study}

In this section, first, we describe real-life black-box DNN's API's that we used to evaluate \prname and then present an overview of \prname's performance.

\subsection{Image-Analyzer}
\label{subsec1}
Image-Analyzer\footnote{https://www.guidancesoftware.com/app/Image-Analyzer-Day-Free-Trial} provides commercial  API  to forensics and 
computer security companies\footnote{https://ecom.image-analyzer.com/knowledgebase.php}. 
Image analyzer API is used for automating flagging offensive content in image forensics, evidence gathering, email and web content. When an image is uploaded to advertised API endpoint it returns a JSON response with probability scores for each category as described below. 

\begin{itemize}
  \item \textbf{Terrorism Category} - Images containing terrorist militants, beheadings, executions, propaganda, acts of terrorism, flags and insignia.
  \item \textbf{Weapons Category} - Images containing handheld weapons such as rifles, machine guns, handguns, knives, swords, grenade launchers and people holding firearms.
   \item \textbf{Pornography Category} - Images that contain commercial pornography, amateur pornography, sexting, selfies, nudity, sex acts, greyscale pornographic images, sexually explicit cartoons and manga.
   \item \textbf{Gore Category} - Images containing graphic violence, bloody wounds, accident victims, Beatings, mutilation, decapitation and other images that contain blood and guts
    \item \textbf{Drugs Category } - Images containing illegal drugs, drug use, drug paraphernalia, plants and symbols relating to drugs
 
\end{itemize}

It is possible for an image to return high probabilities for multiple
categories. There are 24 error codes produced by remote API (Table \ref{errorcodes}).
The higher the probability for a particular category the more likely the image composition matches that category. The default recommended threshold is 50\%, this will provide high detection with false positives of approximately 1-2\%. Increasing the threshold will decrease the false positives and detection rate. Lowering the threshold will increase the false positives and detection rate.

\subsection{Threat Model}
\label{threat_model}
\begin{itemize}
\item Rate Limiting - Image analyzer has a query limiting for test accounts of maximum 10,000 queries per user.
\item Rejection rate - Image analyzer has a strong rejection rate for known black-box attacks as seen by sensitivity prober.
\item Model updates - As advertised in API documentation, models are updated frequently with new training data.
\item Errors - System error -25 in table \ref{errorcodes} can alert the system administrator of attacker activity.
\end{itemize}

The goal of the adversary is to bypass the image content detection engine and achieve zero detection. Bypassing the engine here means for a given image in normal settings probability score of the image is greater than the threshold when compared to attack image with similar content the probability score is lesser than the threshold. In an attacker aware setting system, administrators or maintainers will review both alerts and errors raised by the system. In order to achieve zero detection, the adversary should not only raise any alerts but also errors in the system.

\begin{table}[h!]

\centering
\begin{tabular}{|c| c| c |} 
 \hline
 Error Code  & Type  & Description \\ [0.5ex] 
 \hline\hline

-7 & Runtime & Invalid Command.  \\
\hline
-9 & Runtime  & Invalid Image Data  \\
\hline
-10 & Runtime & Invalid Category Id/s  \\
\hline
-11 &System & Cannot decode the image  \\
\hline
-12 &System & Unable to load Image Reader  \\
\hline
-15  & System  & Unexpected error, please try again.  \\
\hline
-16 & System & Error as returned by the scanning functionality  \\
\hline
-17 & Runtime & Invalid / Empty image URL  \\
\hline
-20 & Runtime  & Database Error  \\
\hline
-21 &Runtime & Empty Image data from URL  \\
\hline
-22 & System & Failed to load IASSL library  \\
\hline
-23 & System & Error in daemon service  \\
\hline
-25  & System & Error by background service.  \\
\hline
-26  & Runtime & Invalid JSON.  \\ [1ex] 

 \hline
\end{tabular}
\caption{System and Run time error codes returned by Image Analyzer API}
\label{errorcodes}
\end{table}

\subsection{Results}

\begin{table}[h!]

\centering
\begin{tabular}{|p{3cm}|p{5cm}|}
 \hline
 ATF Component  & Parameters  \\ [0.5ex] 
 \hline\hline
 Model Bank & Inception v3\cite{szegedy2016rethinking}, Resnet 50,
Resnet 101, Resnet 152\cite{he2015deep}, VGG 16\cite{vgg16},
Adv. trained  Inception v3, Adv.ens. trained Resnet 50\cite{models}   \\
\hline
Sensitivity Prober & Gaussian Noise, JPEG Compression, Random Noise  \\
\hline 
Error Recorder  & -25 System Error  \\
\hline 
Rate Limiter   & \textless 1000 API calls  \\
\hline 
Attack Selector    &  FGSM, I-FGSM, C\&W, JSMA   \\
\hline 
Attack Generator    &  FGSM- $\epsilon$ = 16/256, I-FGSM-20 Iteration, C\&W and, JSMA (Default values from cleverhans library)     \\
\hline 
Feature Influence Calculator   &  Integrated Gradients with 300 Iterations, threshold- 3  \\
[1ex] 
\hline

\end{tabular}

\caption{Different parameters used in ATF for attacking Image Analyzer API }
\label{atf_img}
\end{table}

\begin{table*}[!htb]
\centering
\begin{tabular}{|c|c|c|c|c|c|}
\hline
           & \small FGM  &  \small I-FGM &\small  C\&W & \small JSMA  & \small ATF \\ \hline
\small ResNet-152 & 4\% & 5\%       & 7.5\%         & 3.5\%  & 97\%     \\ \hline
\small ResNet-101 & 3.54\% & 6.5\%       & 10.2\%             & 4.1\%   & 95\%    \\ \hline
\small ResNet-50 & 13\% & 17\%       &  9 \%           & 5\%   & 94 \%    \\ \hline
\small VGG-16     & 10.2\% & 12.5\%       & 14\%           & 12\%     & 84 \%  \\ \hline
\small Adv.Train Inception V3 & 5\% & 10\%       & 8.2 \%            & 11\%  & 96\%       \\ \hline
\small InceptionV3  & 2\% & 7\%       & ErrorCode-25           & 13\%       & 97\% \\ \hline
\small Ensemble Trained ResNet-50  & ErrorCode-25 & ErrorCode-25      & ErrorCode-25            & ErrorCode-25   & 94\%   \\ \hline

\end{tabular}
\caption{Different Attack Images generated using the model bank. The Index column equal model name which is used to generate attack images The cell $(i, j)$ indicates the percent of images out of 400 which bypassed detection for particular attack and model combination. }
\label{model-results}
\end{table*}

\begin{figure}[h!]
    \includegraphics[width=\columnwidth]{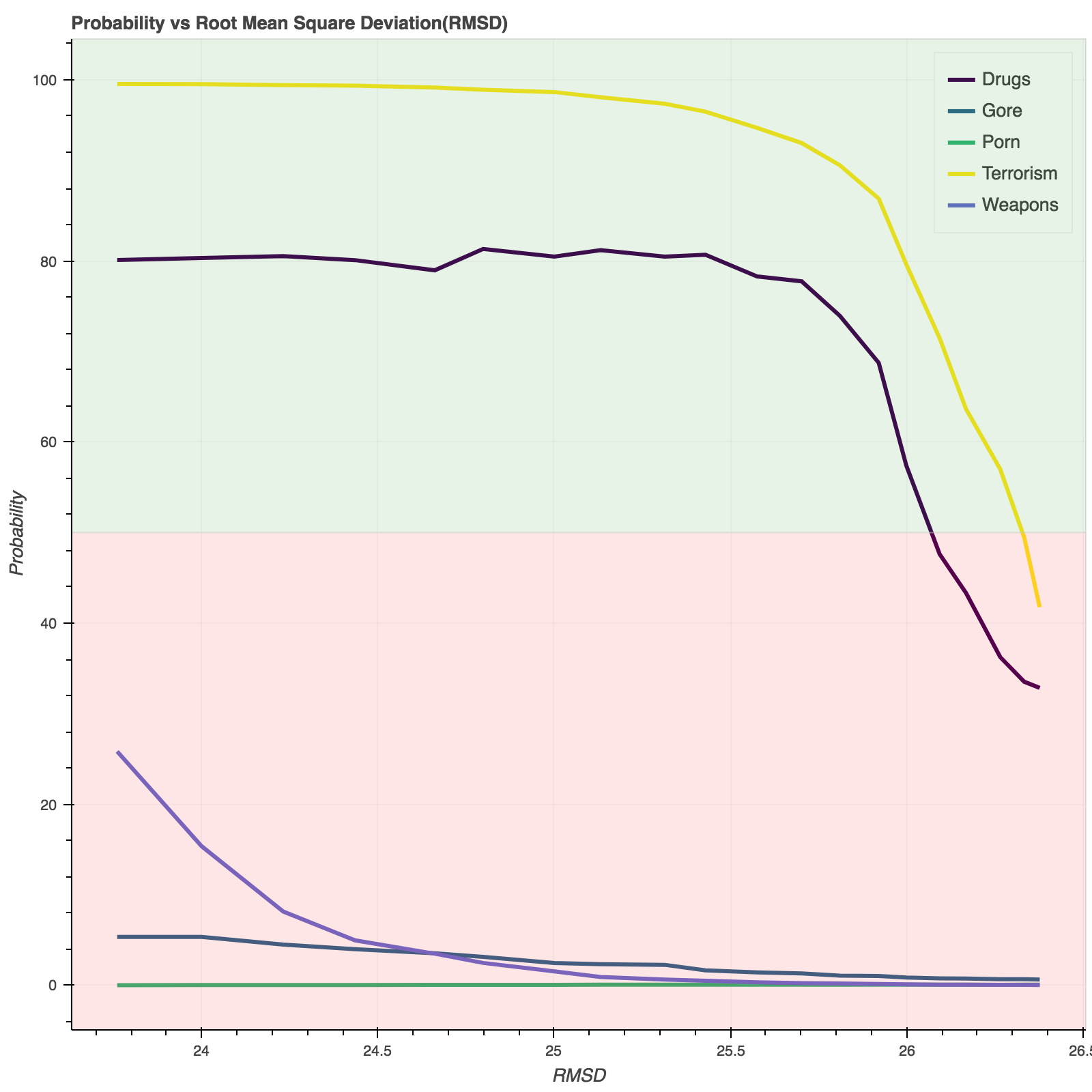}
    \caption{Probability scores of each category  vs CDF of Root Mean Square Deviation, Green and Red  regions indicate detection and bypassing regions respectively. Adversary has to stay in red region to avoid detection.Threshold is set to default value(50)}
    \label{fig:graph1}
\end{figure}

\begin{table*}
    \centering
    \begin{tabular}{c|ccc}\toprule
    Query-based attack  & Attack success & No. of queries & \% of samples generated Error(-25)    \\ \midrule

    Iter. Finite Diff. & 18 & \textgreater 10000 & 72 \\
    Iter. Gradient Estimation  & 55  & 7600 & 45 \\
     ATF  & 97  & 856 & 3 \\

 \bottomrule
    \end{tabular}
    \caption{Comparison of untargeted query-based black-box attack methods. All results are for attacks using the  400 samples from the Imagenet ''\textit{assault rifle, assault gun}'' dataset on  and with an $L_{\infty}$  and perturbation of $\epsilon=32$.}
    \label{tab: query_bb}
\end{table*}

We evaluated Image analyzer API with 400 images with label ''\textit{assault rifle, assault gun}'', these images were randomly drawn from train data of Imagenet.
Due to sensitive nature of categories like terrorism, we did not test the robustness of API w.r.t. other categories.
Table \ref{atf_img} summarizes different parameters used in Adversary Testing Framework. 
As shown in Table \ref{model-results}  attack images generated by different models failed to bypass the detection system. The reasons remote API was able to detect these attacks is either it is trained to detect these published attacks or attack was not transferable with default parameters. In future, we want to explore with parameter tuning instead of default parameters.

We also observed that attack images raised ''backend service error''. This behavior suggests that model detects adversarial samples but does not recover true class of the image. When random noise image is submitted to API we observe no error generated, suggesting that model is sensitive to noise. As noted in Figure \ref{fig:graph1}   the probability for each category drops as distortion (RMSD) increases. Table \ref{real-app} summarizes probability changes of top category of an image w.r.t. to selectively changing pixels using integrated gradient method.
We compare ATF with iterative finite differences and iterative gradient estimation  attacks\cite{2017arXiv171209491} non-targeted black-box attacks in Table \ref{tab: query_bb}, ATF outperforms other attacks in constraint settings.

\begin{table}[h!]

  \begin{tabular}{ | c | c |  }
    \hline
   Image & Prob of Top category,Perturbation \%  \\ \hline
   
    \includegraphics[width=20mm, height=20mm]{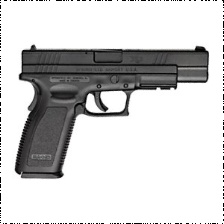} & Weapons:99.99715, 2.2263 \\ \hline

    \includegraphics[width=20mm, height=20mm]{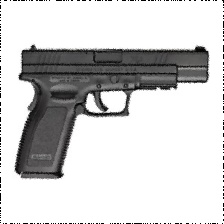} & Weapons:98.24705, 3.6469  \\ \hline

    \includegraphics[width=20mm, height=20mm]{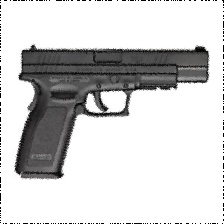} & Weapons:36.24675, 3.7806  \\ \hline
 \includegraphics[width=20mm, height=20mm]{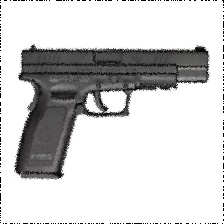} & Weapons:0.133139357611, 4.3276 \\ \hline

  \end{tabular}
  \caption{Images , 
probabilities returned by image analyzer API for Top category and Percent of Perturbing Influence Pixels.}

\label{real-app}
\end{table}

\section{Discussion}

The popularity of DNN  has led to an explosion in the number of DNN libraries, frameworks, and services. A service provider can use in-house infrastructure, data to train a DNN model or outsource model creation to a cloud service such as  Amazon ML\footnote{https://aws.amazon.com/amazon-ai/} and AutoML\footnote{https://cloud.google.com/automl/}. Services developed on top of these models are available to end users for automating tasks which may be security critical.

Non Expert users have no insights into how these models were trained, implemented and how trustworthy are the platforms hosting them. Interpreting outputs of DNN are nontrivial due to the complexity involved. We argue that any framework which tests the security robustness of a black-box DNN has to answer following questions. 

\begin{table*}[t]

\begin{centering}
\begin{tabular}{|p{1in}|p{4in}|}
\hline
{\bf Characteristics of DNN} & {\bf Implication} \\
\hline
Non-transparency & The structure of a trained ML model is difficult for a human to understand. Is the architecture, weights,
and implementation of the model open and
reproducible.\\
\hline
Training scenarios & Does the data set has
enough representation of all classes. Is the
training set poisoned? Does the training set
have inherited bias? \\
\hline 
Error rate & A trained ML model does not guarantee correctness and has a known estimated error rate. Are the real error rates model
experiences close to development process estimate? Does the error rate increase over
time?\\
\hline
Instability & The structure of a trained ML model is emergent and may change arbitrarily even for small changes in training. How much stable outputs are when
the training set changes?\\
\hline
Privacy \& Anti-Discrimination & Is model resistant to Information
leak and fair w.r.t.
predictions? \\
\hline 
\end{tabular}

\caption{Characteristics of machine learning based implementations.}
\label{t:charml}
\end{centering}

\end{table*} 

As a future work, we want to extend ATF to provide the testing methodology for answering each question summarized in Table \ref{t:charml}

\section{Conclusion and Future work}
\label{sec5}

In this paper, we designed and implemented \prname, a domain-independent framework for systematically testing security robustness of black-box DNN with varying threat models and adversary goals. We introduced a new method, Feature Influence Calculator, for bypassing black-box DNN. we tested \prname with a  DNN tool used in digital forensics operating in constraint settings. 
As a future work, we want to extend the \prname to test the robustness of DNN algorithms used in other security domains mainly malware classification, mobile authentication, mobile cloud forensics\cite{Faheem15}, anomaly detection, on-chip DNN models and closed source software which are relevant for Digital Forensics community. We are going to analyze the impact of the use of DNN tools on ISO 27041, 27042 standards and also propose recommendations to these standards to accommodate technologies like DNN.

\end{document}